\documentclass{aa}
\pdfoutput=1
\usepackage{graphicx}
\usepackage{txfonts}
\usepackage{amsfonts,amsmath,amssymb}
\usepackage{bm}
\usepackage{array}
\newcounter{rowno}
\usepackage{threeparttablex}
\usepackage[colorlinks=true,allcolors=blue]{hyperref}
\def\mone{{\rm M}_1}
\def\mtwo{{\rm M}_2}

\def\aone{{\rm a}_1}
\def\atwo{{\rm a}_2}

\def\xeff{\chi_{\rm eff}}
\def\nsamp{{\rm N}_{\rm samp}}

\def\nmrg{{\rm N}_{\rm mrg}}

\def\zmax{z_{\rm max}}

\def\rate{\mathcal R}

\def\perygm{{\rm~yr}^{-1}{\rm Gpc}^{-3}{\rm M}_\odot^{-1}}
\def\peryg{{\rm~yr}^{-1}{\rm Gpc}^{-3}}

\def\mcllow{M_{\rm cl,low}}
\def\mclhigh{M_{\rm cl,high}}

\newcommand{\Ms}{\ensuremath{{\rm M}_{\odot}}}

\newcommand{\eg}{{\it e.g.}}

\newcommand{\ie}{{\it i.e.}}

\newcommand{\beq}{\begin{equation}}
\newcommand{\eeq}{\end{equation}}

\newcommand{\mchirp}{\ensuremath{M_{\rm chirp}}}

\newcommand{\kmps}{\ensuremath{{\rm~km~s}^{-1}}}
\newcommand{\peryr}{\ensuremath{{\rm~yr}^{-1}}}

\newcommand{\mcl}{\ensuremath{M_{\rm cl}}}
\newcommand{\rh}{\ensuremath{r_{\rm h}}}

\newcommand{\tmrg}{\ensuremath{t_{\rm mrg}}}

\newcommand{\nmrgin}{\ensuremath{N_{\rm mrg,in}}}
\newcommand{\nmrgout}{\ensuremath{N_{\rm mrg,out}}}

\newcommand{\nbseven}{{\tt NBODY7}}

\newcommand{\bse}{{\tt BSE}}

\newcommand{\archain}{{\tt ARCHAIN}}

\newcommand{\tevol}{\ensuremath{T_{\rm evol}}}

\newcommand{\fbin}{\ensuremath{f_{\rm bin}}}
\newcommand{\nbin}{\ensuremath{N_{\rm bin}}}
\newcommand{\fobin}{\ensuremath{f_{\rm Obin}}}

\newcommand{\mbh}{\ensuremath{M_{\rm BH}}}

\newcommand{\mrem}{\ensuremath{M_{\rm rem}}}

\newcommand{\vesc}{{\rm v}_{\rm esc}}

\newcommand{\fmrg}{\ensuremath{f_{\rm mrg}}}
\newcommand{\ftz}{\ensuremath{f_{\rm TZ}}}

\begin{document}

\title{Binary black hole mergers from young massive clusters
in the pair-instability supernova mass gap}
\titlerunning{PSN-gap BBH mergers from young massive clusters}
\author{Sambaran Banerjee
	\thanks{Corresponding author. E-mail: sambaran@astro.uni-bonn.de (he/him/his)}
	\inst{1,2}
       }
\institute{
        Helmholtz-Instituts f\"ur Strahlen- und Kernphysik,
        Nussallee 14-16, D-53115 Bonn, Germany
        \and
        Argelander-Institut f\"ur Astronomie,
        Auf dem H\"ugel 71, D-53121, Bonn, Germany
         }
\abstract
{
The recent discovery of the binary black hole (BBH) merger event GW190521, between
two black holes (BHs) of $\approx100\Ms$, and as well as other massive BBH merger events
involving BHs within the pair-instability supernova (PSN) mass gap
have sparked widespread debate on the origin of such extreme gravitational-wave (GW)
events. GW190521 simultaneously triggers two critical questions: how BHs can
appear within the `forbidden' PSN gap and, if so, how they get to participate in
general-relativistic (GR) mergers?
}
{
In this study, I investigate whether dynamical interactions in young massive clusters (YMCs)  
serve as a viable scenario for assembling PSN-gap BBH mergers.
}
{
To that end, I explore a grid of 40 new evolutionary models of a representative YMC of
initial mass and size $\mcl=7.5\times10^4\Ms$ ($N\approx1.28\times10^5$) and $\rh=2$ pc, respectively.
The model grid ranges over metallicity $0.0002 \leq Z \leq 0.02$
and comprises initial cluster configurations of King central concentration parameters
$W_0=7$ and 9. In each model, all BH progenitor stars are initially in primordial
binaries following observationally-motivated distributions. All cluster models
are evolved with the direct, relativistic N-body code $\nbseven$ incorporating up to date remnant formation,
BH natal spin, and GR merger recoil schemes.
}
{
BBH mergers from these model cluster computations agree well with the masses
and effective spin parameters, $\xeff$, of the events from the latest gravitational-wave transient catalogue
(GWTC). In particular,
GW190521-like, \ie, $\approx200\Ms$, low $\xeff$  events are produced via dynamical merger
among BHs derived from star-star merger products.
GW190403\_051519-like, \ie, PSN-gap, highly asymmetric, high $\xeff$ events result from
mergers involving BHs that are spun up via matter accretion or
binary interaction. The resulting present-day, differential intrinsic merger rate density,
within the PSN gap, well accommodates that from GWTC.
}
{
This study demonstrates that, subject to model uncertainties, the tandem of massive binary evolution
and dynamical interactions
in $\lesssim100$ Myr-old, low metallicity YMCs in the Universe
can plausibly produce GR mergers involving PSN-gap BHs
and in rates consistent with that from to-date GW observations.
Such clusters can produce extreme events alike GW190521 and GW190403\_051519.
The upper limit of the models' GW190521-type event rate is within the
corresponding LIGO-Virgo-KAGRA (LVK)-estimated rate limits, although the typical model rate
lies below the LVK's lower limit. The present YMC models yield a merger rate
density of $0-3.8\times10^{-2}\peryg$ for GW190521-type events. They produce GW190403\_051519-like
events at a rate within $0-1.6\times10^{-1}\peryg$ and their total BBH-merger yield
within the PSN gap is $0-8.4\times10^{-1}\peryg$. 
}
\maketitle

\section{Introduction}\label{intro}

Within just a few years after the first detection \citep{2016PhRvL.116f1102A} of gravitational wave (hereafter GW)
from a merging binary black hole (hereafter BBH), we are already approaching a `golden era'
of GW and multi-messenger astronomy \citep{Branchesi_2016,Mapelli_2018,Meszaros_2019,Mandel_2021}.
Recently,
the LIGO-Virgo-KAGRA collaboration \citep[hereafter LVK;][]{Asai_2015,Acernese_2015,KAGRA_2020}
has published,
in their GW transient catalogue (hereafter GWTC)
\footnote{\url{https://www.gw-openscience.org/eventapi/html/GWTC/}},
about 90 candidates
of general relativistic (hereafter GR) compact binary merger events from
until the end of their third observing run, `O3'. The up to date GWTC
includes all event candidates
from the LVK's first, second (`O1', `O2'; \citealt{Abbott_GWTC1}),
and third (`O3'; \citealt{Abbott_GWTC2,Abbott_GWTC3}) observing runs
including those in their `Deep Extended Catalogue' \citep{Abbott_GWTC2.1}.

GWTC contains several events that have triggered
widespread debate regarding the events' origin or formation channel. 
Perhaps the most explored event, since its announcement, is
GW190521 \citep{Abbott_GW190521}: a merger of two black holes (hereafter BHs)
of estimated masses $\mone=95.3_{-18.9}^{+28.7}\Ms$ and $\mtwo=69.0_{-23.1}^{+22.7}\Ms$
and effective aligned spin parameter $\xeff=0.03_{-0.39}^{+0.32}$
(\citealt{Abbott_GWTC2}; the limits correspond to 90\% credibility).
$\xeff$ \citep{Ajith_2011} is a measure of the spin-orbit alignment of
a merging binary and is defined as
\begin{equation}
\xeff \equiv \frac{\mone\aone\cos\theta_1 + \mtwo\atwo\cos\theta_2}{\mone+\mtwo}
	= \frac{\aone\cos\theta_1 +  q\atwo\cos\theta_2}{1 + q}.
\label{eq:xdef}
\end{equation}
Here, the GR-inspiralling masses $\mone$, $\mtwo$, with mass ratio
$q\equiv\mtwo/\mone$, have, respectively,
Kerr vectors $\vec\aone$, $\vec\atwo$ that project with angles
$\theta_1$, $\theta_2$ on the orbital angular momentum vector just
before the merger.
The Kerr vector or Kerr parameter
(also addressed as dimensionless spin vector or dimensionless spin parameter),
$\vec a$, is defined as \citep{Kerr_1963}
\begin{equation}
\vec a = \frac{c\vec S_{\rm BH}}{G\mbh^2}
\end{equation}
where $\vec S_{\rm BH}$ is the total angular momentum vector of a Kerr BH of (non-spinning) mass
$\mbh$.

Another notable PSN-gap event is GW190403\_051519 (hereafter GW190403)
which is lighter than GW190521 but, in contrast, is
highly mass-asymmetric ($\mone=88.0_{-32.9}^{+28.2}\Ms,\mtwo=22.1_{-9.0}^{+23.8}\Ms$)
and spin-orbit aligned ($\xeff=0.70_{-0.27}^{+0.15}$).
GW190426\_190642 has masses and $\xeff$ similar to GW190521
although this event is of much higher false alarm rate (hereafter FAR), 4.1 $\peryr$,
as opposed to ${\rm FAR}=2\times10^{-4}\peryr$ for GW190521.
The current GWTC
includes several additional candidates of BBH mergers with the primary
being within the PSN gap.

The excitement is natural since
both components of this BBH merger lie well within the `forbidden'
pair-instability supernova (hereafter PSN) mass gap between
$45\Ms\lesssim\mrem\lesssim120\Ms$ \citep{Belczynski_2016a,Woosley_2017}.
Starting to evolve from their zero age main sequence (hereafter ZAMS), massive single
stars are not expected to produce any compact remnant over this remnant-mass range
and, instead, explode completely due to PSN occurring in their late
evolutionary stages \citep{Langer_2007,Woosley_2017,Mapelli_2020}.
The lower limit of the PSN gap occurs due pulsation pair-instability supernova (hereafter PPSN)
that episodically sheds the hydrogen envelope of the parent star until
a helium core of $\approx45\Ms$ is retained \citep{Woosley_2017,Woosley_2020} which then directly
collapses into a BH. The upper limit of the gap results when
the progenitor star becomes massive enough that PPSN and PSN quenches
(see, \eg, \citealt{Ziegler_2021})
and the evolved star can directly collapse into an intermediate mass BH (hereafter IMBH). 
See Fig.~2 of \citet{Banerjee_2020} for the `standard' ZAMS mass-final mass relation,
exhibiting the PSN gap at different metallicities.

Therefore, the observation of BHs within the PSN gap, at least at a first glance,
is a signature of an outlier BH population \citep{Baibhav_2021,OBrien_2021}
and intrigues more exotic scenarios of BH formation and merger.
Several scenarios have already been invoked to explain GW190521-type events:
primordial BHs \citep[\eg,][]{Clesse_2017,Carr_2018,DeLuca_2019},
BHs derived from Population III stars \citep[\eg,][]{Tanikawa_2021,Tanikawa_2021b,Ziegler_2021},
gas accretion onto BHs in dense proto-clusters \citep[\eg,][]{Roupas_2019},
GR coalescences in field hierarchical systems \citep[\eg,][]{Fragione_2020,VignaGomez_2021,Hamers_2021},
and stellar collisions and dynamical interactions inside dense stellar clusters.

In this study, a young massive cluster (hereafter YMC) origin of PSN-gap BBH mergers
is investigated. Dynamical interactions among stellar-remnant BHs inside dense star
clusters is an intriguing scenario for generating PSN-gap and other types
of massive BBH mergers
\citep[][]{Perna_2019,ArcaSedda_2020b,Baibhav_2020,Banerjee_2020c,Rizzuto_2021,Rizzuto_2021b,Gerosa_2021,Mapelli_2020b}
since the scenario naturally allows for prolonged post-processing
of BHs and BBHs formed as a result of massive binary evolution.
Dynamical interactions also enable star-star and star-remnant mergers  
of kinds which would not be possible with the evolution of isolated massive binaries
alone \citep[][]{DiCarlo_2020,Gonzalez_2020,Rastello_2021}.
Such interactions would potentially lead to `forbidden' BHs and mergers involving them, depending on the
cluster's properties and its stellar and multiplicity content. Since the most
massive members in a star cluster undergo dynamical processing the earliest,
the most massive BBH mergers are expected to occur early in the cluster's evolution \citep{Banerjee_2017}.
As shown in earlier works \citep{Banerjee_2010,Banerjee_2017,Rodriguez_2018},
dynamical BBH mergers in a YMC can occur inside the cluster due to
hierarchical-system interactions (\eg, Kozai--Lidov oscillation, chaotic triple-interaction;
\citealt{Kozai_1962,Katz_2011,Lithwick_2011})
or close flyby interactions, within the centrally segregated BH subsystem \citep{Banerjee_2010,Morscher_2015}.
A fraction of the BBHs (and other types of compact binaries) ejected from the cluster, either dynamically
or due to natal kick, can also merge within the Hubble time \citep{1993Natur.364..423S,PortegiesZwart_2000}.

It should, however, be borne in mind that it is possible that the `standard' PSN mass gap (see above)
is simply
an artefact of the `standard' theoretical stellar evolution models.
The PPSN mass cap can potentially be much higher or
the PSN mass gap can even be nonexistent.
This is supported by both
stellar evolutionary models \citep[][]{Belczynski_2020b,Farmer_2020,Woosley_2021,Vink_2021,Costa_2021}
and the observed population of merging BHs \citep[][]{Abbott_GWTC3_prop,Edelman_2021}. 
In that case, GW190521-like events can simply be explained by the
standard isolated binary evolution scenario \citep{Belczynski_2020d}.
The present study bases on the existence of the standard PSN gap and
involves a stellar-remnant model that exhibits such a gap.

This paper is organized as follows.
Sec.~\ref{compute} describes a new set of direct N-body computations of model YMCs.
Sec.~\ref{result} discusses the outcomes, focussing on PSN-gap mergers. Sec.~\ref{GW190521_like}
describes the inferred rate of GW190521-like events and Sec.~\ref{GW190403_like} covers
that of GW190403-like mergers. Sec.~\ref{summary} summarizes the findings and
discusses the implications of various model assumptions and the limitations. 

\begin{figure*}[!ht]
\centering
\includegraphics[width = 18.0 cm, angle=0.0]{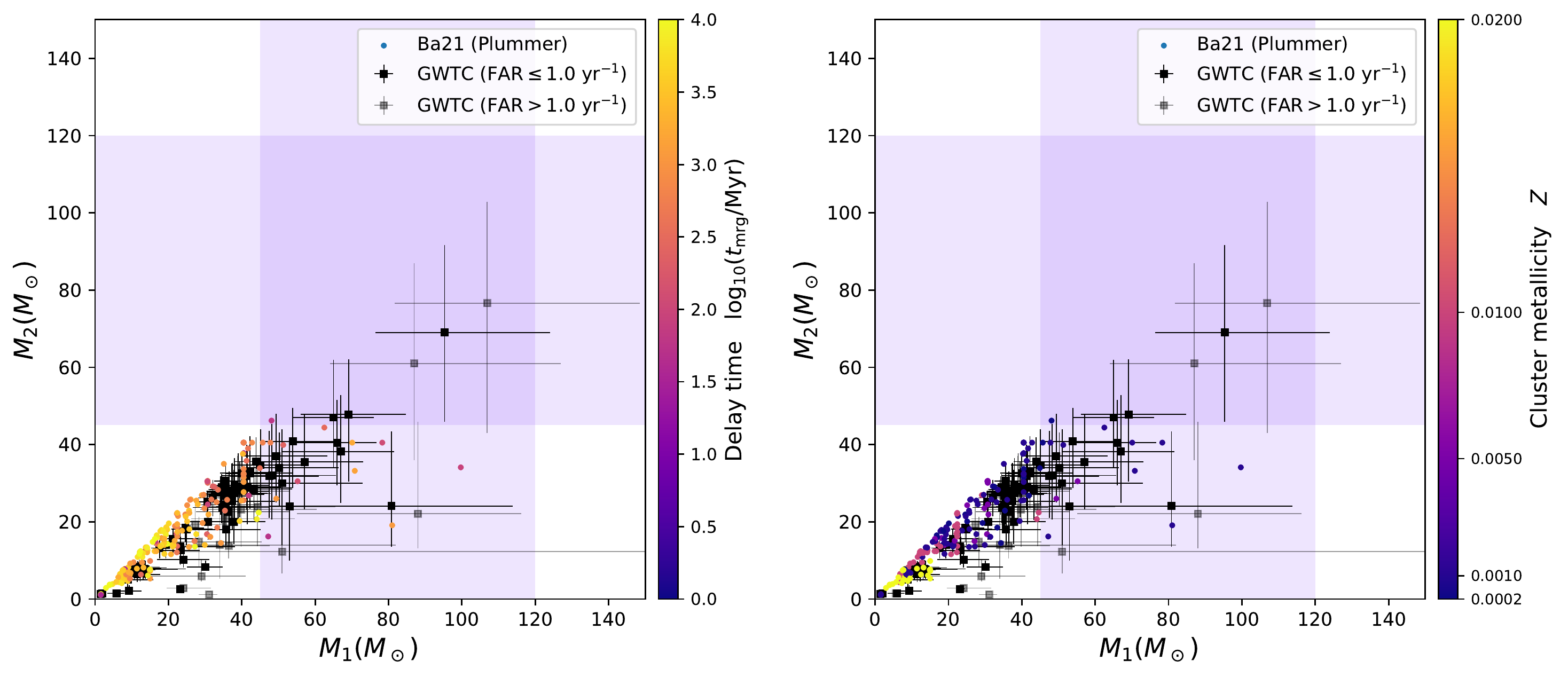}
\includegraphics[width = 18.0 cm, angle=0.0]{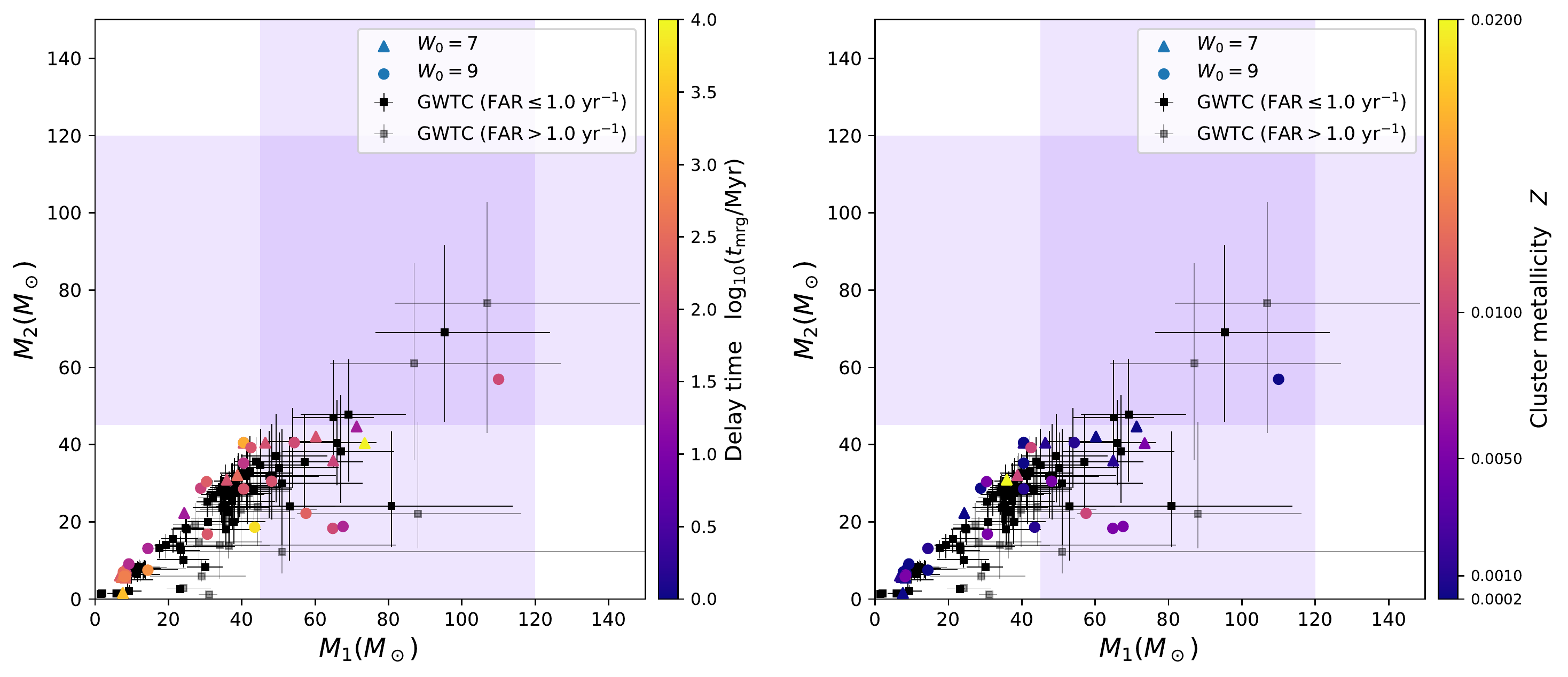}
\caption{{\bf Top panels:} primary mass, $\mone$, versus secondary mass, $\mtwo$ ($\mone\geq\mtwo$),
of all BBH mergers
from the computed model star clusters of Ba21 (filled circles). These data points are colour-coded
according to the merger delay time, $\tmrg$ (left panel), and the parent model cluster's
metallicity, $Z$ (right panel). The GWTC events' data points (black, filled squares)
and their corresponding
90\% credible intervals (horizontal and vertical error bars) are indicated, where
events with FAR $\leq1\peryr$ and $>1\peryr$
are distinguished by different shades. The blue-shaded
regions over $45\Ms-120\Ms$ on each panel represent the PSN mass gap.
(The blue shadings are placed in the background so that they do not
influence the colours of the data points.)
{\bf Bottom panels:} the same as the top panels except that the mergers from the
present computed clusters (Table~\ref{tab_nbrun}) with King concentration parameters
$W_0=7.0$ (filled triangles) and 9.0 (filled circles) are shown.
}
\label{fig:m1m2}
\end{figure*}

\section{Computations}\label{compute}

Close dynamical interactions, star-star mergers, and star-remnant mergers are generally favoured
in massive star clusters having profiles of high central concentration. This is due to the resulting high
central stellar density and efficient mass segregation. Therefore, massive BH formation
and their mergers can be expected to preferably occur in such systems. Since massive, PSN-gap
BH formation and their involvement in GR mergers is the focus of this study, massive,
concentrated model clusters are specifically considered.
In this work, model star clusters with initial mass of $\mcl=7.5\times10^4\Ms$ 
(initial number of stars $N\approx1.28\times10^5$) are taken to be representatives of YMCs. Such a
cluster mass is comparable to those of most massive Galactic and local-group YMCs. It is
also representative of moderate mass `super star clusters' \citep{PortegiesZwart_2010}.
The stellar content of
these star cluster models is motivated by the cluster models described in
\citet[][hereafter Ba21]{Banerjee_2020c}.
The initial density and kinematic profiles
of the clusters follow the \citet{King_1966} model with half-mass radius $\rh=2$ pc
and King dimensionless potential of $W_0=7$ and $9$. Such size is typical for YMCs
\citep{PortegiesZwart_2010,Krumholz_2019}.

Previous results in Ba21 and \citet{Banerjee_2021}, where the models use the \citet{Plummer_1911}
initial conditions, \ie, with the central concentration comparable to $W_0\approx5-6$ \citep{Heggie_2003},
suggest that massive clusters ($\mcl\geq5\times10^4\Ms$) are capable of producing and merging PSN-gap BHs.
An objective of the present study is to see how somewhat more centrally
peaked initial models fair in this which is why $W_0=7$ and 9 are chosen here (less
concentrated model would be uninteresting in this regard due to the reasons given
above). The
King profile offers the freedom to alter the central concentration without
altering the cluster's overall length scale (as measured by $\rh$).

Another objective is to see how pc-scale clusters fair as opposed to highly
concentrated, sub-pc-scale clusters of $\rh$ comparable to widths of molecular-cloud
filaments, the latter conditions having been recently studied by several authors
\citep{DiCarlo_2020,Rastello_2021,Rizzuto_2021b}.
(Near) monolithic, (near) gas-free, pc-scale YMCs are something that we do observe and can measure
with relative unambiguity in various surveys \citep{Mathieu_2000,PortegiesZwart_2010,Kuhn_2013,Krumholz_2019}.
The gas-free nature of such clusters implies that any potential violent-relaxation phase
and its impact on the cluster \citep{Banerjee_2018b} can be ignored and the cluster can be assumed to evolve
solely via secular dynamical evolution, as is the case with the present model clusters.

As in Ba21, the initial models are composed of
ZAMS stars of masses $0.08\Ms\leq m_\ast\leq150.0\Ms$
that are distributed according
to the canonical initial mass function \citep[hereafter IMF;][]{Kroupa_2001}
and have an overall (see below) primordial-binary fraction of $\fbin=5$\%.  
However, as in Ba21, the initial binary fraction of the
O-type stars ($m_\ast\geq16.0\Ms$), which are initially paired only among themselves,
is taken to be $\fobin(0)=100$\%, to be consistent with the observed high binary fraction
among O-stars in young clusters and associations
\citep[\eg,][]{Sana_2011,Sana_2013,Moe_2017}.
The O-star binaries are taken to initially follow
the observed orbital-period distribution of \citet{Sana_2011}
and a uniform mass-ratio distribution.
The initial orbital periods of the non-O-star primordial binaries follow the period
distribution of \citet{Duq_1991} and their mass-ratio distribution is also uniform.
The initial eccentricity of the O-star
binaries follows the \citet{Sana_2011} eccentricity distribution and that
for the rest of the binaries obeys the thermal eccentricity distribution
\citep{Spitzer_1987}. As explained in \citet{Banerjee_2017b}, such a scheme for including
primordial binaries provides a reasonable compromise between the economy of computing
and consistencies with observations.

The model clusters are evolved with the star-by-star, direct N-body evolution code $\nbseven$ \citep{Aarseth_2012},
which is updated in several aspects as detailed in \citet[][hereafter Ba20]{Banerjee_2020} and Ba21.
These updates mainly enable up-to-date stellar wind mass loss and remnant formation in the
code and as well implement numerical relativity (hereafter NR)-based GR merger recoil
and spin recycling of merged BHs.
The latter allows for on-the-fly and consistent treatment of BBH merger products.
The primary `engine' for stellar and binary evolution in
$\nbseven$ is $\bse$ \citep{Hurley_2000,Hurley_2002} and that for post-Newtonian
(hereafter PN) evolution of compact binaries and higher order systems is $\archain$ \citep{Mikkola_1999,Mikkola_2008}. 
In the present computed models, the `F12-rapid+B16-PPSN/PSN' (see Ba20) remnant-mass
prescription is applied. For single star evolution, this remnant prescription allows for the formation of
stellar remnants, of mass $\mrem$, maintaining a neutron star (NS)-BH mass gap between
$2\Ms\lesssim\mrem\lesssim5\Ms$ \citep{Fryer_2012}
and a PSN mass gap between $45\Ms\lesssim\mrem\lesssim120\Ms$
(\citealt{Belczynski_2016a,Woosley_2017}; see also Fig.~2 of Ba20).

In the present computations, zero Kerr parameter is assigned to
all BHs derived from single stars or from members of non-mass-transferring or non-interacting binaries,
as per the BH-formation models of \citet{Fuller_2019a} - the FM19 BH natal spin model.
If, after formation, a BH undergoes matter accretion due to a BH-star merger or mass transfer in a binary
or if a BH is born in a tidally-interacting (or symbiotic) binary,
its Kerr parameter is set to the maximally spinning value of $a=1$. In the event of a BH-star merger
(the formation of a BH Thorne--Zytkow object, \citealt{TZ_1975} or BH-TZO),
$\ftz=0.95$ fraction of the merging star's mass
is assumed to be accreted onto the BH. Also, in star-star mergers, $\fmrg=0.2$ fraction of the secondary's
mass ($\leq10$\% of the total merging stellar mass) is assumed to be lost in the merger process
\citep{Gaburov_2008,Glebbeek_2009}.
In the present runs, the NR treatment is updated
to the more recent GW-recoil formulae of \citet{Lousto_2012} and final-spin formulae of \citet{Hofmann_2016}.  

For each $W_0$ of the initial King profile, five metallicities are taken, namely, $Z=0.0002$, 0.001, 0.005,
0.01, and 0.02.
Four random model realizations of each ($W_0$, $Z$) pair,
which are subjected to a solar-neighborhood-like external field
\footnote{As seen in recent studies such as \citet{Webb_2021},
the dynamical evolution of YMCs similar to the present models remains
practically unaffected by large alterations to the galactocentric distance.},
are evolved for 300 Myr.
These 40 newly computed models are listed in Table~\ref{tab_nbrun}.
%

\begin{figure*}[!h]
\centering
\includegraphics[width = 18.5 cm, angle=0.0]{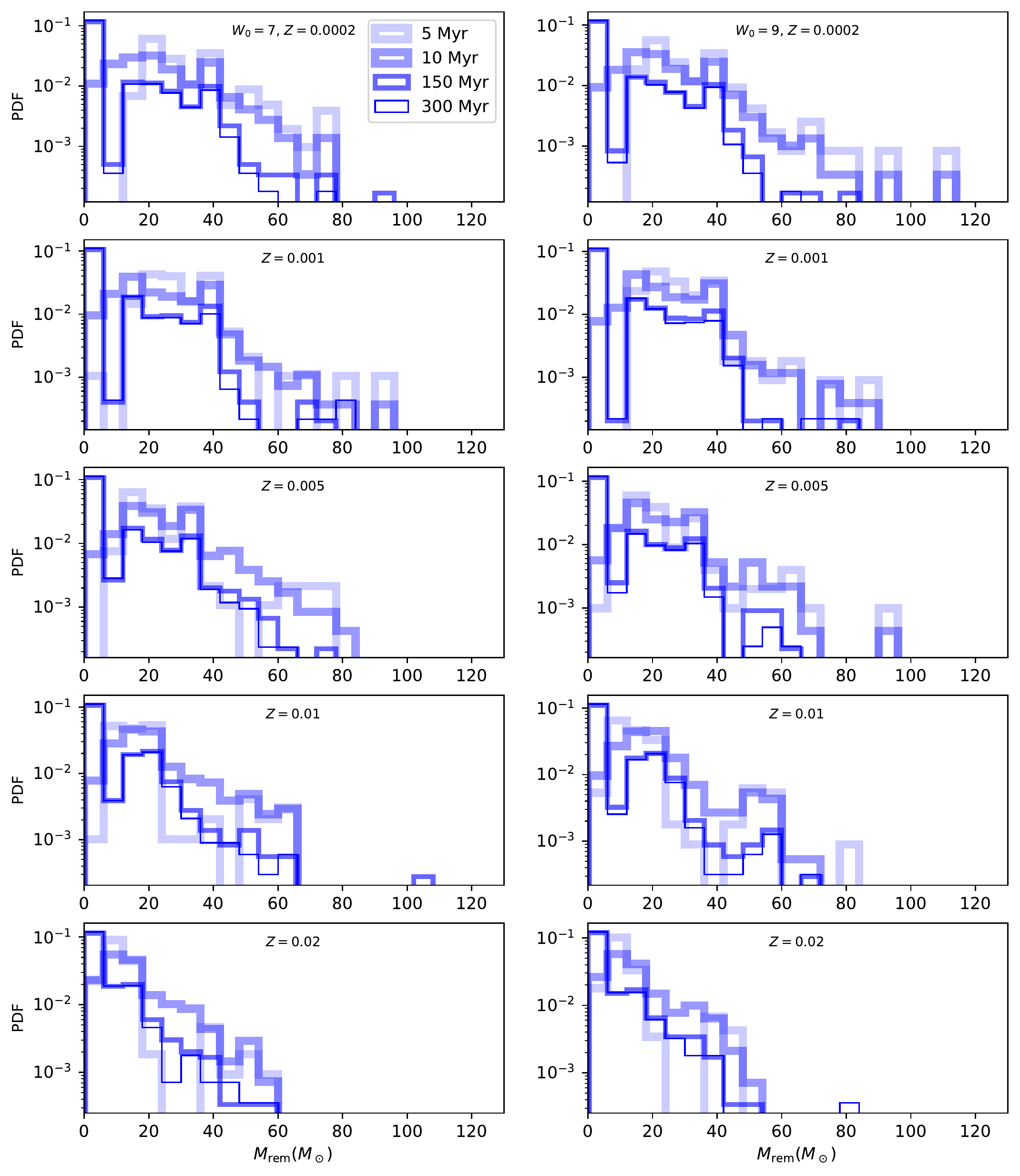}
\caption{The mass distributions of stellar remnants remaining
in the computed model clusters (Table~\ref{tab_nbrun})
at 5 Myr, 10 Myr, 150 Myr, and 300 Myr cluster-evolutionary times (legend). The distributions, at a given time, from
all (4) models for a particular $W_0$ and $Z$ are combined in one histogram. The panels in the left (right)
column correspond to the models with $W_0=7$ ($W_0=9$) with $Z$ as indicated in each panel's title.}
\label{fig:remdist_W07}
\end{figure*}

\begin{figure*}
\centering
\includegraphics[width = 19.0 cm, angle=0.0]{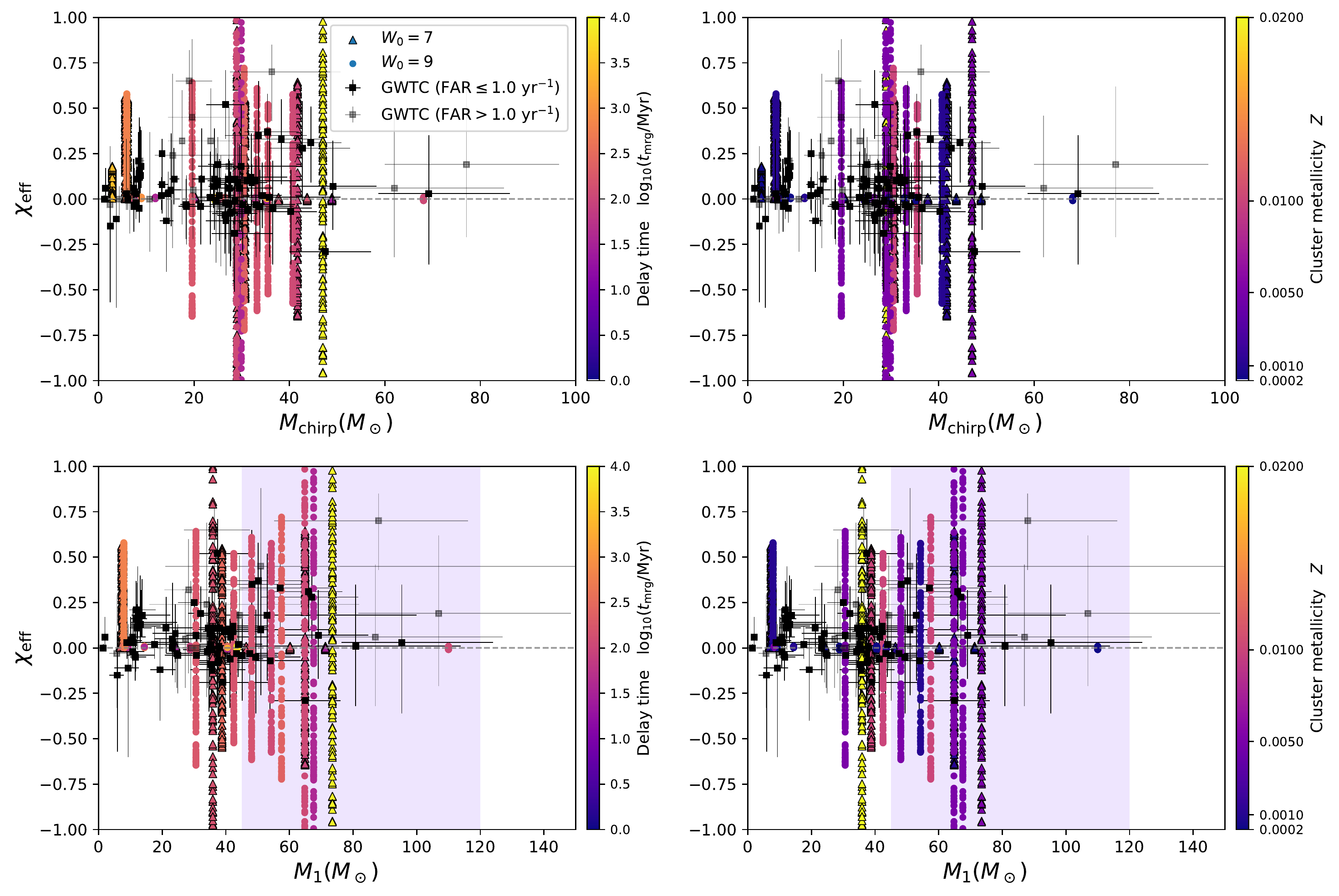}
\caption{Effective aligned spin parameter, $\xeff$, versus chirp mass, $\mchirp$ (top
panels), and primary mass, $\mone$ (bottom panels), of all BBH mergers from the
computed model clusters of Table~\ref{tab_nbrun}. For each merger, $\xeff$ values for 
a number of random orientations of the merging BHs' spins
are plotted (Sec.~\ref{result}). The legends are the same as in Fig.~\ref{fig:m1m2}
except that a black edge is applied to the filled triangle symbol for improved visibility.}
\label{fig:xeff_mass}
\end{figure*}

\begin{figure*}
\centering
\includegraphics[width = 19.0 cm, angle=0.0]{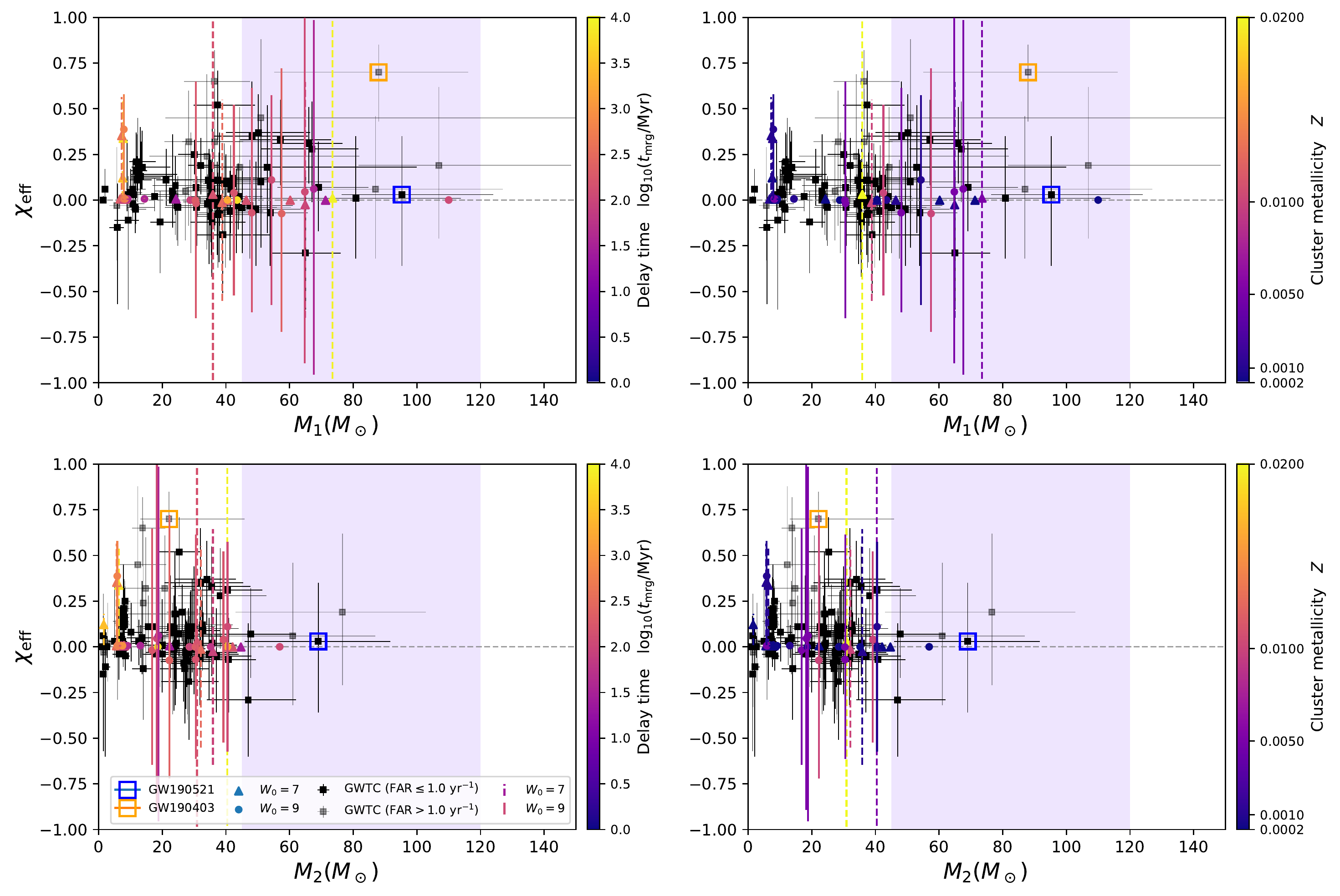}
\caption{Effective aligned spin parameter, $\xeff$, versus primary mass, $\mone$ (top
panels), and secondary mass, $\mtwo$ (bottom panels), of all BBH mergers from the
computed model clusters of Table~\ref{tab_nbrun}. For each model merger,
$\xeff$ values for a number of random orientations of the merging BHs' spins
are obtained (Sec.~\ref{result}) which are shown as a data point (mean value)
with a vertical error bar (range). The legends of the data points
are the same as in Fig.~\ref{fig:m1m2}.
The error bars (in dashed line and solid line for mergers from the $W_0=7$ and $W_0=9$ models, respectively)
share the same colour code as the corresponding data points.
The LVK events GW190521 and GW190403 are highlighted with two empty squares (which
are coloured differently for improved legibility and do not correspond to the
color code).
}
\label{fig:xeff_mass2}
\end{figure*}

\section{Results}\label{result}

Fig.~\ref{fig:m1m2} (bottom panels) show the primary mass, $\mone$, versus secondary mass,
$\mtwo$ ($\mone\geq\mtwo$), of the compact-binary mergers from the computed models.
All these events are in-cluster or ejected BBH mergers, except for one that is an NS-BH merger.
The BBH-merger masses are, overall, well consistent with those from GWTC
(black, filled squares with error bars; events with FAR $\leq1\peryr$ and $>1\peryr$
are distinguished by different shades).
The NS-BH merger is also
consistent with the LVK-detected NS-BH merger events
\citep{Abbott_NSBH_2021},
as seen near the bottom left corners of Fig.~\ref{fig:m1m2} (bottom panels).
For comparison, analogous plots are obtained from the computed models of
Ba21 (Fig.~\ref{fig:m1m2}; top panels) whose mergers show similarly
good agreement with the LVK events. 

The presently computed cluster models are more efficient in producing
PSN-gap BBH mergers: the 40 models produce similar number of mergers with $\mone$ and/or $\mtwo$ within
the gap as the 65 models of Ba21. This is due to
the more centrally-peaked initial profiles of the present
models (Sec.~\ref{compute}) than those of
Ba21 (despite similar $\rh$ for both model sets).
However, in both sets, mergers involving a PSN-gap BH happen only in models
with $Z\leq0.005$ (Fig.~\ref{fig:m1m2}).
In the present models, BHs in the PSN gap appear due to (i) star-star
mergers and (ii) BH-TZO accretion. Depending on the stellar-evolutionary
age of the merging stars, a star-star merger can result in an over-massive H-envelope
of the merged star while its He-core mass being below the PPSN/PSN threshold.
With sufficient H-envelope mass, such a merged star would evolve into a
direct-collapse BH within the PSN-gap \citep[\eg,][]{Spera_2019,Banerjee_2020,DiCarlo_2020b}.
Such a remnant mass range is `forbidden'
for single stars evolving directly from ZAMS. The high, 95\% BH-TZO accretion, as adopted here,
would also push BHs into the PSN gap as a result of sufficiently massive BH-star mergers.

In the Ba21 models, a third channel has also produced PSN-gap BBH mergers in the model
runs with FM19 (vanishing) BH natal spins, namely, second generation (hereafter 2G) mergers
\footnote{In this work, BHs derived directly from the nuclear evolution
of stellar progenitors will be called first generation or 1G BHs. If a 1G BH
gains mass due to stellar matter accretion, it will still be 1G. A 1G+1G BBH merger
will result in a 2G BH, a BBH merger involving at least one 2G BH will
yield a 3G BH, and so on. This convention is often followed in the literature.}.
In those models,
most 2G BBH mergers have happened with delay times $\tmrg>300$ Myr.
In the present models, although a few 2G BHs remain in the clusters
until the end of the runs at 300 Myr evolutionary time (the majority of the
2G BHs being ejected from their host clusters at their formation
due to GW recoil kick; see Sec.~\ref{compute}), none of them could
get involved in another merger by this time as can be expected based
on the Ba21 models.

The most massive, GW190521-like merger obtained from the present model set
(Fig.~\ref{fig:m1m2}; bottom panels) is a 1G, in-cluster merger between two PSN-gap
BHs, both of which are derived from star-star merger products. The merger happened
in one of the $Z=0.0002$, $W_0=9$ models. Fig.~\ref{fig:remdist_W07}
plots the mass distributions of stellar remnants retained within the cluster
at different evolutionary times, for all the 40 computed models. To improve counts
over the mass bins and as well facilitate comprehension, distributions, at a given cluster age,
from all the four models with a given $W_0$ and $Z$ are combined in one histogram.
The figure demonstrates the overall trend that the maximum mass of the BHs
retained in a cluster at a given age increases with decreasing $Z$ and increasing $W_0$,
as can be expected.

\begin{figure*}
\centering
\includegraphics[width = 18.0 cm, angle=0.0]{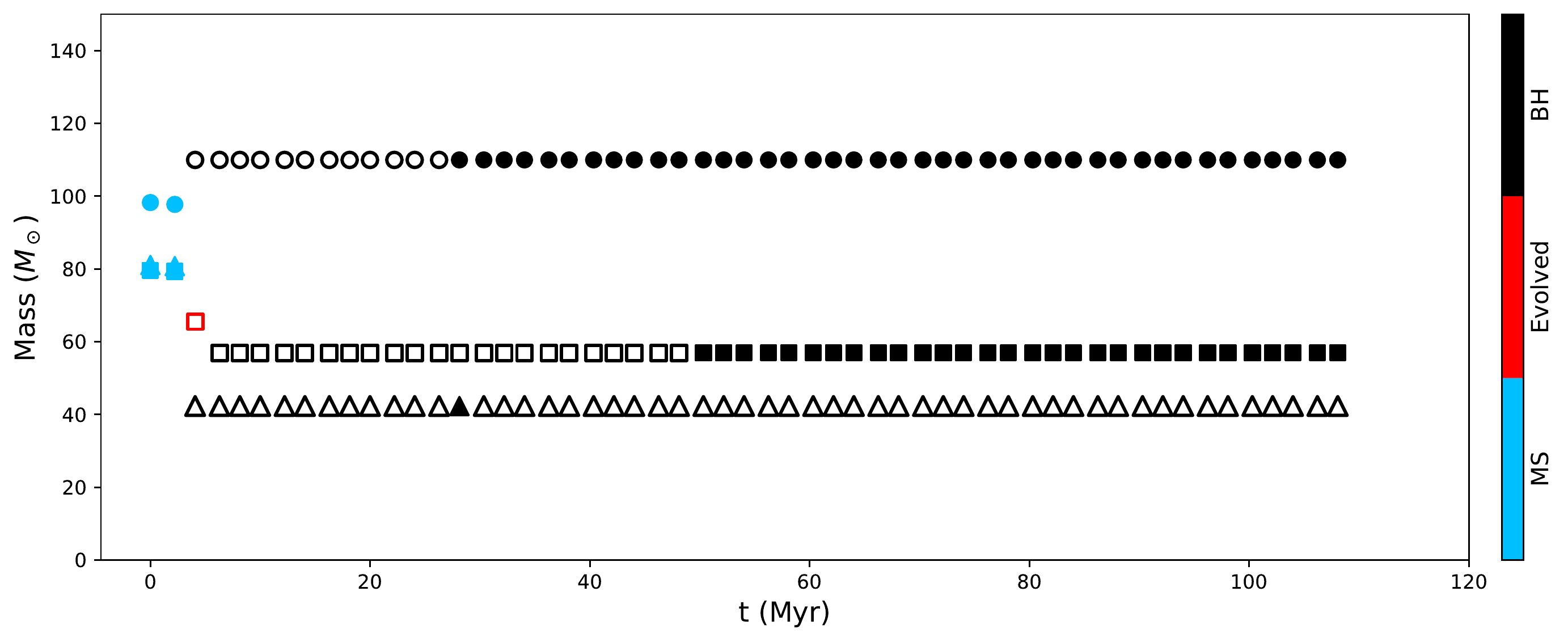}
\includegraphics[width = 18.0 cm, angle=0.0]{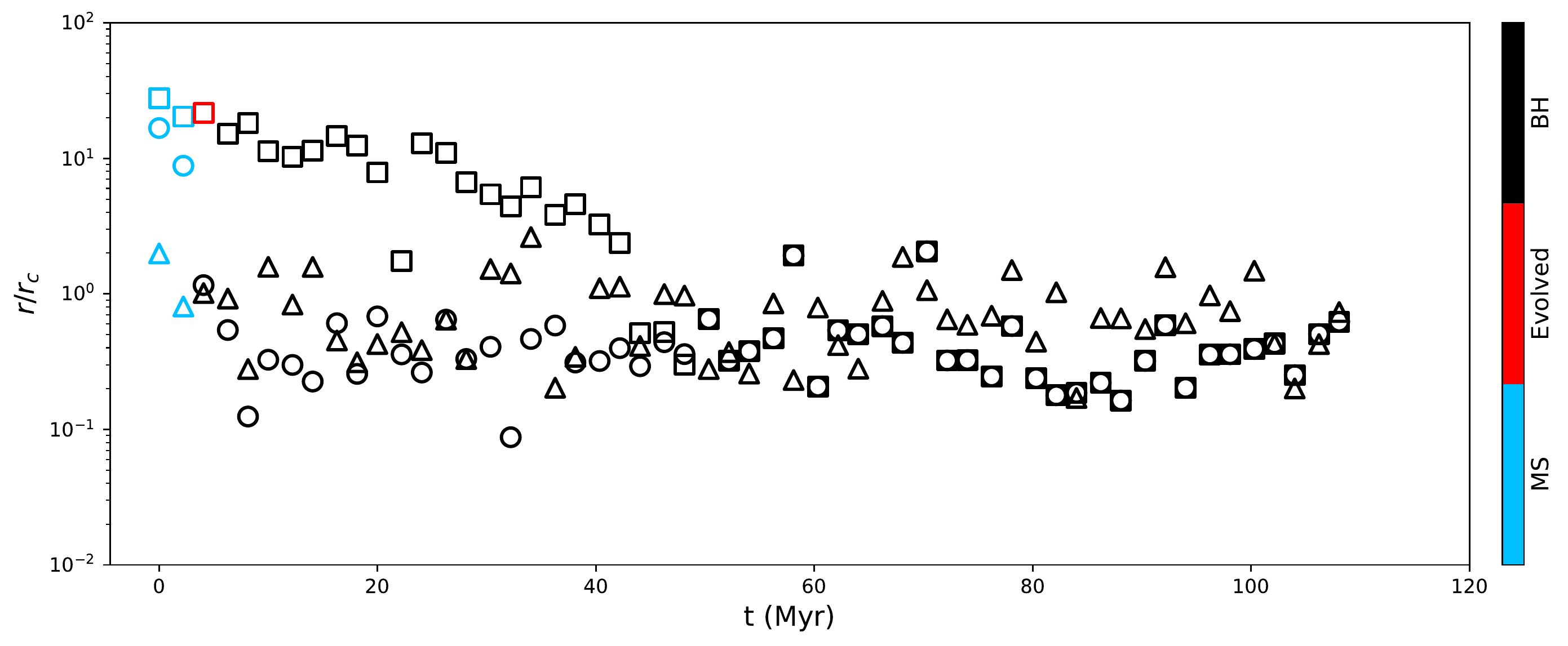}
\caption{The timeline of the GW190521-like merger event that occurred inside one of the computed
	cluster models with $W_0=9$, $Z=0.0002$. In both panels the evolutions for the three cluster members,
	which are most relevant for the event, are shown from the start of the computation until the
	merger. The top panel shows the mass evolution of these objects: the trace of the merger primary and secondary
	mass is indicated by a circle and a square, respectively, and that of the third `perturber'
	is represented by a triangle. The symbols are color-coded according to the stellar-evolutionary
	stage (main sequence, MS or beyond main sequence, Evolved or BH).
	If an object is a member of a
	binary then its symbol is filled else it is plotted empty.
	The bottom panel traces the radial position of these objects in units of the cluster's
	instantaneous core radius $r_c$, $r_c$ being determined from the cluster's stellar distribution
	by applying the \citet{Casertano_1985} method.
	The time evolution of $r_c$, 10\% Lagrangian radius, and half mass radius
	of all model clusters (Table~\ref{tab_nbrun}) are shown in Fig.~\ref{fig:sizeevol}.
	The same symbol shapes and color coding as in the top panel are used in the bottom panel:
	for better visibility of overlapping symbols, the symbol filling is not applied in the bottom panel.
	Both the primary and the secondary
	BHs are born form single stars, thereby having vanishing natal spins (Sec.~\ref{compute}). Both the
	progenitor single stars are merged primordial binaries: the BH can be more massive than the
	plotted ZAMS progenitor depending on the mass of the latter's companion, the
	age of the primordial binary's merger, and the mass loss during the merger (Sec.~\ref{compute}).
	Although formed within the cluster well separated, the primary and secondary BHs segregate
	to the cluster center and pair up dynamically from $\approx50$ Myr. The third BH becomes
	bound to this BBH dynamically only a few Myr before the merger (lower panel). The
	BBH merger happens due to this triple-interaction \citep{Samsing_2017,Banerjee_2018,Banerjee_2020c}.}
\label{fig:event190521}
\end{figure*}

\begin{figure*}
\centering
\includegraphics[width = 14.5 cm, angle=0.0]{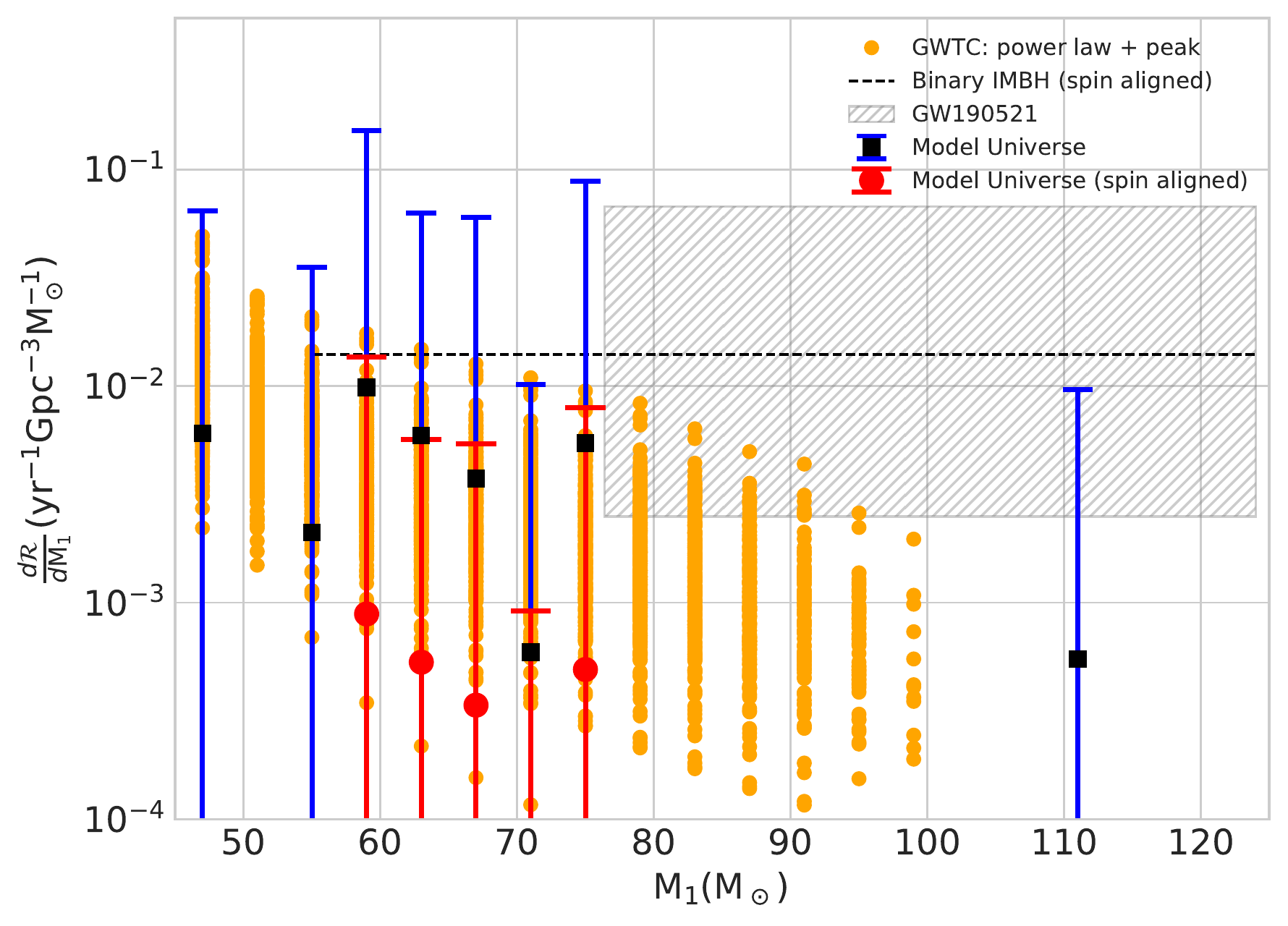}
\caption{The present-day, differential intrinsic BBH merger rate density
(black-filled squares with blue error bars),
from the Model Universe (Sec.~\ref{GW190521_like}) composed of the computed model clusters of
Table~\ref{tab_nbrun}, as a function of
merger primary mass, $\mone$, within the PSN mass gap
(here taken to be $45\Ms \leq \mone \leq 120\Ms$).
At each $\mone$-bin, the average rate over the model sets (Sec.~\ref{GW190521_like})
is indicated with the black-filled square and the corresponding vertical, blue
error bar indicates the maximum and minimum rates for the bin.
The orange dots are random draws (300 per bin) of the posteriors of BBH differential
intrinsic merger rate density, over the same range of $\mone$, as obtained by
the LVK \citep[][their power law + peak model]{Abbott_GWTC3_prop}.
The hashed region spans over the 90\% credible intervals of the intrinsic merger rate density
of GW190521-like events and GW190521's primary mass, as estimated by the LVK
\citep{Abbott_GW190521,GW190521_rate}.
The horizontal, black-dashed
line indicates the LVK-estimated upper limit of the
merger rate density of equal mass, aligned spin ($\xeff>0.8$)
BBHs of total mass $200\Ms$ \citep{GW190521_rate}. The line spans
from the lower primary mass limit of GW190403 to the upper primary mass limit of GW190521.
The red-filled circles and the corresponding vertical, red error bars
represent, respectively, the average and the limits of the aligned spin fractions of the Model Universe
differential BBH merger rate density over $55\Ms\lesssim\mone\lesssim75.0\Ms$.
The LVK-estimated total rates are divided by the bin width ($4\Ms$), to represent
them in the $d\rate/d\mone-\mone$ plane. The lower limit of the Model Universe rate is zero in each bin.
Due to the logarithmic scale along the vertical axis,
rate values below $10^{-4}\perygm$ are not shown in this figure.}
\label{fig:diffrate_PSNgap}
\end{figure*}

\begin{figure*}
\centering
\includegraphics[width = 18.0 cm, angle=0.0]{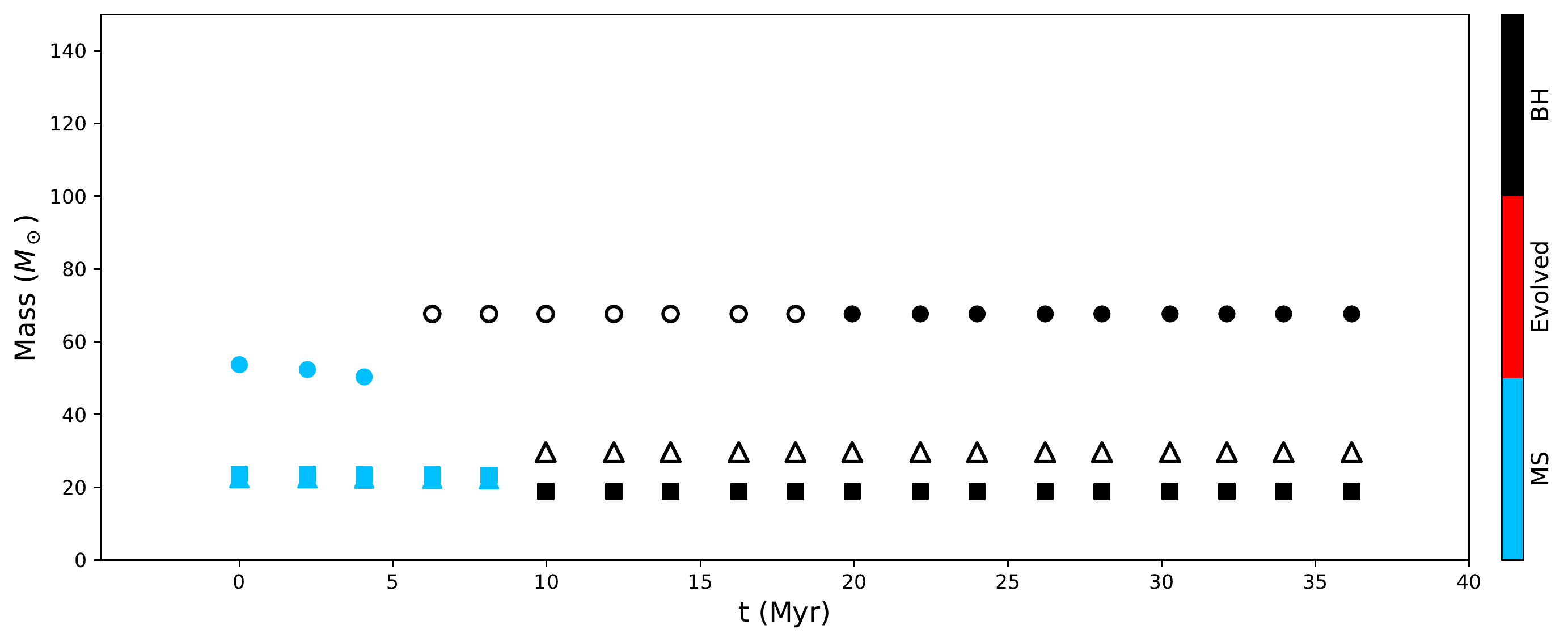}
\includegraphics[width = 18.0 cm, angle=0.0]{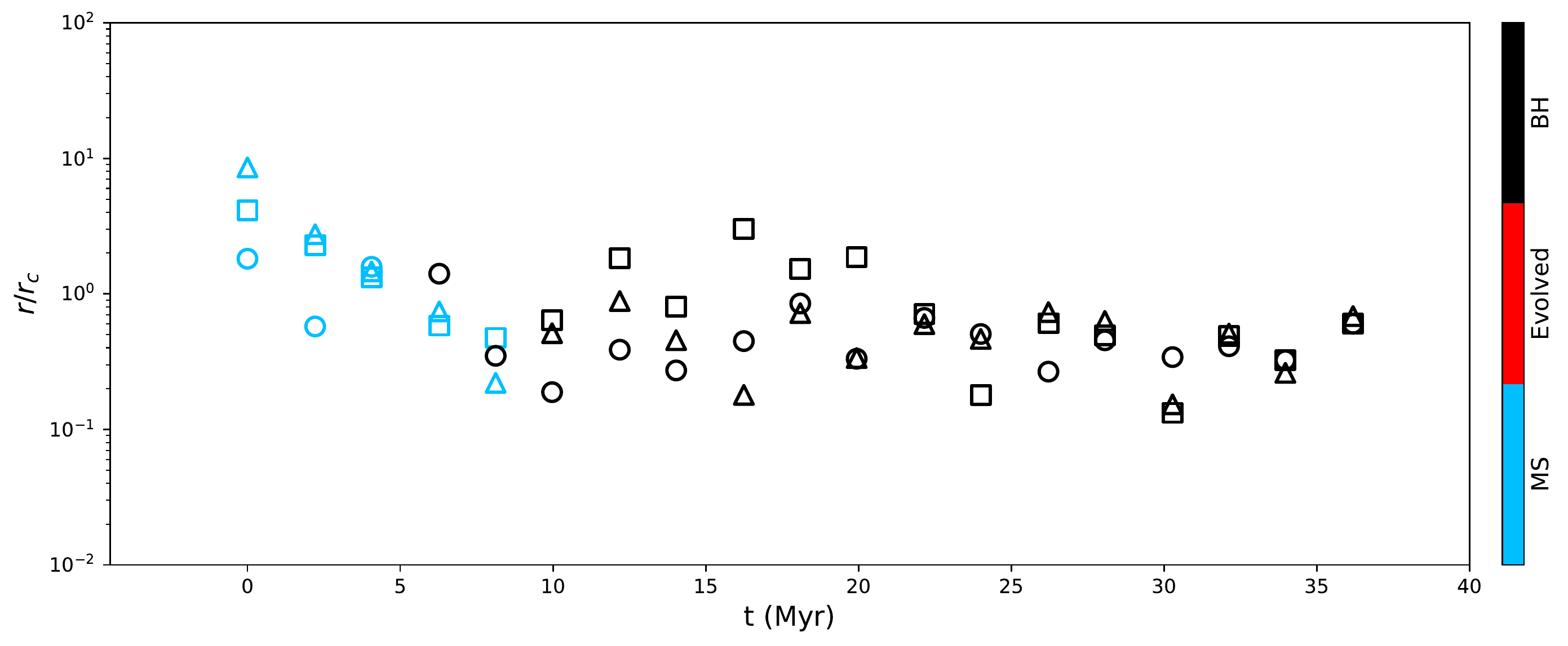}
\caption{The timeline of a GW190403-like merger event that occurred inside one of the computed
	cluster models with $W_0=9$, $Z=0.005$. The meanings of the symbols are the same as in
	Fig.~\ref{fig:event190521}. Here, early in the cluster evolution, the merger primary BH
	is born in a primordial binary and 
	gains mass via BH-TZO accretion from its stellar companion. The merger secondary BH is born
	in a symbiotic binary. This is how both of the merging BHs become maximally spinning ($a_1=a_2=1$),
	given the present scheme of assigning BH spins (Sec.~\ref{compute}). As in the example
	of Fig.~\ref{fig:event190521}, the BBH merger happens due to dynamical pairing of the merging BHs
	and, thereafter, a triple-interaction involving a third perturber BH. In this case, both
	interactions happen only a few Myr before the merger (lower panel). All other GW190403-type
	mergers in the present models are also in-cluster dynamical mergers involving at least one
	such `spun-up' BH, except one which is an ejected merger (see Fig.~\ref{fig:xeff_mass}).}
\label{fig:event190403}
\end{figure*}

Fig.~\ref{fig:xeff_mass} plots the effective (aligned) spin parameter, $\xeff$, of the GR mergers
from the present models against their chirp mass, $\mchirp$ (top panels), and
primary mass (bottom panels). A merging system's chirp mass is defined as  
\begin{equation}
\mchirp \equiv \frac{(\mone\mtwo)^{3/5}}{(\mone+\mtwo)^{1/5}}.
\label{eq:mchirp}
\end{equation}
For every GR merger from the computed models, a range of potential
values of $\xeff$ is plotted in Fig.~\ref{fig:xeff_mass}
by assigning independent, random values to
$\theta_1$ and $\theta_2$ in Eqn.~\ref{eq:xdef}. If the components of the merging
binary are derived from stellar progenitors that were members
of different primordial binaries or were initially single stars,
\ie, if the merging binary is dynamically assembled
then the merging members are uncorrelated.
Accordingly, $(\theta_1, \theta_2)$ are taken to be isotropically
oriented and values over the range $0\leq(\theta_1,\theta_2)\leq2\pi$
are assigned to them. On the other hand, if the merging components
are derived from stellar progenitors that were
members of the same primordial binary then they would, plausibly, be at least
partially correlated. Therefore, in this case,
values over the range $0\leq(\theta_1,\theta_2)\leq\pi/2$
are assigned. For every merger from the models, the corresponding
values of $\mone$, $\mtwo$, $\aone$, $\atwo$ are taken
directly from the model
\footnote{Special outputs are arranged in the updated $\nbseven$
for recording $\mone$, $\mtwo$, $\aone$, $\atwo$ of the in-cluster
and ejected mergers. In Fig.~\ref{fig:xeff_mass}, each merger from
the computed models is plotted with 100 random orientations of
$\theta_1$ and $\theta_2$. For the convenience of plotting,
a small value of $a_i=10^{-2}$ is assigned if $a_i=0$.}.

For comparison, event data from GWTC
are also plotted in Fig.~\ref{fig:xeff_mass}. The LVK data points do
exhibit an overall bias towards positive $\xeff$, at least when the
most probable values (squares) are considered. This is indicative of
contribution from additional channels which tend to produce spin-aligned
mergers, \eg, isolated binary evolution. Mixing with additional
channels will be addressed in a future work (see \citealt{Banerjee_2021b} in this context).  
Nevertheless, considering the 90\% credible limits (error bars) the
computed data points are consistent with the LVK data. In particular,
the computed data well encompass the most massive observed PSN-gap mergers (of $\mone\geq55\Ms$)
with nearly vanishing, mildly aligned, and highly aligned $\xeff$ values.

Fig.~\ref{fig:xeff_mass2} re-plots the model $\xeff$ as mean values with
error bars, against
$\mone$ (upper panels) and $\mtwo$ (lower panels),
and highlights the events GW190521 and GW190403.
Fig.~\ref{fig:xeff_mass2}
shows that the present model clusters produce only one GW190521-like
and 5 GW190403-like events, in the sense that the model mergers' $\mone$,
$\mtwo$, and $\xeff$ all lie within or enter the 90\% credible intervals (the LVK-data error bars)
of the respective parameters of these events of interest.

\subsection{Rate of GW190521-like events}\label{GW190521_like}

The present computed cluster models yield BBH mergers that well resemble
notable PSN-gap events discovered by the LVK, in terms of, both, merging masses ($\mone$, $\mtwo$)
and spin-orbit alignment ($\xeff$). This is evident from
Figs.~\ref{fig:m1m2}, \ref{fig:xeff_mass}, and \ref{fig:xeff_mass2}.
The GW190521-like merger event is an outcome of an
in-cluster, dynamically assembled and triggered merger between
two BHs of $\approx110\Ms$ and $57\Ms$,
in one of the $Z=0.0002$, $W_0=9$ models. Both of these
BHs are formed due to evolution of star-star merger products with over-massive
H-envelope (see above):
both of these stellar mergers happened between components of
massive primordial binaries. Since the BHs are derived
directly from the (merged) stellar progenitors, they have zero
natal spins as per the adopted BH natal spin model (Sec.~\ref{compute}).
After forming as single BHs, they eventually get paired up via exchange
interactions and merge due to a dynamical triple-interaction \citep{Banerjee_2018},
resulting in an event with masses and $\xeff$ well within the 90\%-credible limits of those
of GW190521 and GW190426\_190642. 
The detailed timeline of this event, from the formation of the merging BHs up to
their merger, is depicted in Fig.~\ref{fig:event190521}
based on data directly from the N-body simulation and is described further in the figure's caption.

The merger events from the present computed models are utilized to
evaluate the differential merger rate density within the PSN gap.
This is done by applying the same cluster population synthesis approach
as described in \citet{Banerjee_2021}. The methodology
is detailed in this reference which, therefore, is not repeated here.
In the present application, a computed $\mcl=7.5\times10^4\Ms$
cluster is taken as the representative YMC. Accordingly, when applying
Eqn.~5 of \citet{Banerjee_2021} the YMC birth mass range is taken to be
$[\mcllow, \mclhigh]=[5.7\times10^4\Ms, 10^5\Ms]$.
This gives an average cluster mass of $\langle\mcl\rangle=7.5\times10^4\Ms$
for a power-law cluster mass function of index $\alpha=-2$.

Population syntheses are performed with the merger outcomes from
4 different model sets having $[W_0=7: Z=0.0002, 0.001, 0.005, 0.01, 0.02]$
and 4 different model sets having $[W_0=9: Z=0.0002, 0.001, 0.005, 0.01, 0.02]$
(Table~\ref{tab_nbrun}). With each cluster model set,   
3 independent Model Universe sample cluster populations, each of size $\nsamp=5\times10^5$,
are constructed to obtain 3 values of the present-day event count, $\nmrg$.
These, in turn, yield 6 different differential merger rate density
profiles (a reference and a pessimistic rate for each $\nmrg$;
see the above mentioned study) per model set.
As in the above mentioned study, the `moderate-Z' metallicity-redshift
dependence of \citet{Chruslinska_2019} is applied and
the detector visibility horizon is taken at $\zmax=1.0$.

Fig.~\ref{fig:diffrate_PSNgap} shows the resulting present-day, differential intrinsic merger
rate density profiles, $d\rate/d\mone$, with respect to merger primary mass, within
the PSN gap. Shown are the Model Universe average rates (black-filled squares) and
the upper and lower limits (blue error bars) over the different
model sets, at each $\mone$-bin. As seen, the present-day, PSN-gap intrinsic merger rate density profiles
from the model clusters well accommodate those estimated from GWTC
(\citealt{Abbott_GWTC2_prop}; orange dots). However,
the lower limit of the Model Universe $d\rate/d\mone$ is zero for all the mass bins in
Fig.~\ref{fig:diffrate_PSNgap}.
The upper limit of the Model Universe present-day merger rate density of
GW190521-like events is also well within the LVK-limits of the same
(\citealt{GW190521_rate}; hashed rectangle)
but the corresponding average rate is deficient by several factors.

\subsection{Rate of GW190403-like events}\label{GW190403_like}

The present computed models also produce several BBH mergers of $\mone$, $\mtwo$, and $\xeff$
similar to GW190403, lying well within the event's 90\% confidence limits,
as seen in Figs.~\ref{fig:m1m2}, \ref{fig:xeff_mass}, and \ref{fig:xeff_mass2}.
In such model events, at least one of the merging BHs has $a=1$ due
to undergoing BH-TZO accretion or mass accretion in a binary (Sec.~\ref{compute}), prior
to participating in dynamical pairing and the merger. The high $\xeff$
is due to chance alignment in dynamical pairing. Fig.~\ref{fig:event190403}
depicts the timeline of one of such events that occurred in
one of the $W_0=9$, $Z=0.005$ models.

The red-filled circles with error bars in Fig.~\ref{fig:diffrate_PSNgap} indicate
the Model Universe, spin-aligned differential merger rate density of GW190403-like events, in the
mass range $55\Ms\lesssim\mone\lesssim75\Ms$ over which such events occur
in the computed models. These rates are obtained by multiplying the
full differential rates, over the same $\mone$ range,
with the $p$-value for $\xeff\geq0.8$ of an isotropic $\xeff$ distribution
corresponding to $\aone=\atwo=1$ and $q=0.25$ (GW190403-like mass ratio).
The Model Universe, spin aligned merger
rate remains below the LVK-estimated upper limit of
aligned spin ($\xeff\geq0.8$), equal mass, $200\Ms$ IMBH-IMBH merger rate
(\citealt{GW190521_rate}; black-dashed line).

Note that the above, LVK-reported IMBH-IMBH merger rate density
should be taken only as a reference upper limit in this study. Equal mass
IMBH-IMBH mergers of $200\Ms$ and high $\xeff$ neither occur in the present models
nor has yet been detected by the LVK. The upper merger rate limit is inferred by the LVK based
on artificially injected events \citep{GW190521_rate}. As of now, LVK has not reported a
merger rate that applies specifically to GW190403-like events.
The vast majority of the 2G BHs (which BHs can approach $100\Ms$
and be of high Kerr parameter) leave the clusters right after the
1G-1G BBH mergers that form them, due to the associated GW recoil kick (Sec.~\ref{compute}).
Although
$80\Ms-120\Ms$ star-star-merger-product and mass-accreted BHs remain
in the clusters, they get dynamically ejected within $\lesssim100$ Myr
(Fig.~\ref{fig:remdist_W07}) unless they get engaged
in a massive 1G merger (Sec.~\ref{GW190521_like}).

Table~\ref{tab_rates} lists the Model Universe limits of the present day,
intrinsic merger rate densities of the various event types. The lower limit
of these total rates and as well of the differential rates (see above)
is zero since some of the cluster model sets used in the population
synthesis (Sec.~\ref{GW190521_like}) do not produce any PSN-gap BBH mergers.
The set to set variation of the merger yield also results in the rather
large uncertainty in the Model Universe differential rate profile
within the PSN gap (Fig.~\ref{fig:diffrate_PSNgap}). The rate densities of
GW190403- and GW190521-type events in Table~\ref{tab_rates} incorporate only those
fractions of Model Universe mergers whose $\mone$, $\mtwo$, and $\xeff$ all lie
within the 90\% credible intervals of the respective event parameters.

\begin{table*}
\centering
	\caption{Intrinsic merger rate densities from the computed models.
	}
\label{tab_rates}
\begin{tabular}{cc}
\hline
	Event type & Intrinsic merger rate density$/\peryg$ \\
\hline
	PSN-gap    & 0.0 - $8.4\times10^{-1}$ \\
	GW190521   & 0.0 - $3.8\times10^{-2}$ \\
	GW190403   & 0.0 - $1.6\times10^{-1}$ \\
\hline
\end{tabular}
	\tablefoot{Quoted are the ranges of present-day, intrinsic rate densities of
	Model Universe BBH mergers (right column), that resemble specific observed
	GW events (left column; Fig.~\ref{fig:xeff_mass2}).
	}
\end{table*}

\section{Summary and concluding remarks}\label{summary}

This study investigates dynamical interactions among stars and BHs in YMCs
as a potential mechanism for forming BHs within the PSN mass gap and engaging them
in GR mergers. To that end, evolutionary models of YMCs (Sec.~\ref{compute}, Table~\ref{tab_nbrun})
of representative initial mass and size $7.5\times10^4\Ms$
and 2 pc, respectively, are explored. The model clusters also include an
observationally-motivated population of primordial binaries.
The model evolutions are computed with an updated
version of the direct N-body code $\nbseven$ that incorporates up to date
schemes of stellar mass loss, stellar remnant formation, BH natal spin, and GR-merger recoil.

These computed models produce BBH mergers that are well consistent
with the merger masses ($\mone$, $\mtwo$) of the GWTC events
(Sec.~\ref{result}; Fig.~\ref{fig:m1m2}). This
is as well true for the evolutionary cluster models of Ba21 but the
present models tend to produce BBH merges involving  
PSN-gap BHs more efficiently, owing to their initial configurations
of higher central concentration (Sec.~\ref{compute}, Table~\ref{tab_nbrun}).
A high central concentration accelerates the central segregation of the most
massive cluster members, thereby facilitating star-star and star-BH mergers
which, in turn, facilitate the formation of PSN-gap BHs (Sec.~\ref{result} and references therein).
The model mergers' spin-orbit alignments ($\xeff$) are also consistent with   
the to-date LVK events (Fig.~\ref{fig:xeff_mass}).
In particular, the present models do produce BBH mergers resembling GW190521 and GW190403,
in all aspects ($\mone$, $q$, $\xeff$; Fig.~\ref{fig:xeff_mass2}).

Here, the large, $\approx100\Ms$, component masses and vanishing $\xeff$ of the GW190521-like
event arise due to involvement of BHs derived directly from star-star merger products
whose BHs have zero natal spins (Secs.~\ref{compute}, \ref{GW190521_like}).
In contrast, the high $\xeff$ of the GW190403-like events is due to involvement
of BHs that are spun-up via matter accretion (via mass transfer in a binary or from a BH-TZO)
or being a former member of a tidally-interacting binary (Secs.~\ref{compute}, \ref{GW190403_like}).
The high central concentration
and efficient mass segregation in the model clusters also facilitate the high
mass-asymmetry (see Ba21) in the GW190403-type mergers. The present models
produce mergers of mass ratio down to $\approx0.25$ (similar to Ba21).
However, beyond $\mone\gtrsim30\Ms$, the present
pure dynamical mechanism does not reproduce the bias
towards $\xeff>0$, as apparent in the LVK data (Fig.~\ref{fig:xeff_mass}).
This suggests contribution from additional merger channels; \eg,
isolated evolution of field massive binaries \citep[\eg,][]{Belczynski_2020,Wong_2021,Olejak_2021}.
Further GW events and improved spin measurement in GW observations
will better suggest the channel(s) for BBH mergers of high $\xeff$
magnitudes like GW190403.

The key uncertainties in the cluster model ingredients are the adopted
small star-star merger mass loss and large BH-TZO accretion fractions
($\leq10$\% and 95\%, respectively; see Sec.~\ref{compute}). These
settings optimize massive BH formation in the computed cluster models.
Some support for a $\lesssim10$\%
mass loss in a star-star merger come from `sticky-star'-type
hydrodynamic calculations \citep[\eg,][]{Lombardi_2002,Gaburov_2008}.
The nature of BH-TZO and the resulting matter accretion onto the BH is,
currently, mostly elusive. However, an analogy with the direct collapse
BH formation scenario may provide qualitative justification to the assumed
high BH-TZO accretion \citep{Banerjee_2020b}. Another important
uncertainty is the spin-up of BHs: at present, all BHs undergoing
potential scenarios for gaining angular momentum
(mass transfer from a stellar companion, membership of a symbiotic binary,
BH-TZO accretion) are assigned the maximal spin ($a=1$) irrespective of the BH's
natal spin (Sec.~\ref{compute}). Angular momentum deposition onto a BH is, largely, poorly
understood and the setting of $a=1$ is likely an extremization. A qualitative
basis for this model setting is that the typical spin angular momentum of a star
\citep[\eg,][]{Lang_1992,Hurley_2000}
or orbital angular momentum of a massive-stellar binary
\citep[\eg,][]{Tauris_2017} is
much higher than that of a maximally spinning stellar mass BH
\footnote{Of course, $a=1$ represents the extremal value of
the Kerr parameter and this value is often taken to represent
a highly or near maximally spinning BH. In reality, the exact value
of unity won't be achieved since an increasing specific angular momentum
of the BH would also enhance angular momentum draining via, \eg,
outflow of the accreting matter or GW radiation. However, any value of $a$ infinitesimally close
to unity is feasible and in astrophysical scenarios $a$
can become practically unity \citep[see, \eg,][]{Kesden_2008,Benson_2009}.
Since there is no physical bound on how close to unity $a$ can
get as a BH continues to accrete angular momentum from its surrounding or
companion, $a=1$ can be taken as a representative of a BH that is subjected
to a reservoir of angular momentum.
In practice,
taking $a=1$ or close to $1$ will cause a negligible difference to
the present results \citep[see, \eg,][]{ArcaSedda_2021b}.}
.

The GW-merger recoil of the 2G BHs, as incorporated in the present models
based on recent NR results (Sec.~\ref{compute}),
ejects most of them from these model clusters which have
moderate, $\vesc\approx40\kmps$, central escape speed.
This is why 2G BBH mergers do not contribute to highly aligned, PSN-gap
BBH mergers (GW190403-type events) or any other
GR merger involving a highly spinning BH, in the present YMC models.
(The much larger number of and much longer term evolutionary
cluster models in Ba21 have produced a few
2G BBH mergers.) Note, however, that in galactic nuclear clusters,
where $\vesc\gtrsim100\kmps$, 2G BHs are more likely to engage 
in hierarchical BBH mergers despite their GW recoil \citep{Mahapatra_2021},
which BHs can even be of $\approx100\Ms$ and low spin
\citep{Antonini_2019,Belczynski_2020c}.

The present YMC models, nevertheless,
well accommodate the LVK present-day, differential intrinsic BBH merger rate density
within the PSN gap (Sec.~\ref{GW190521_like}, Fig.~\ref{fig:diffrate_PSNgap}).
The aligned-spin ($\xeff>0.8$) fraction of GW190403-type
mergers, as obtained from these models,
make up a present-day rate density that is consistent with
the LVK-estimated reference upper limit of the rate density of
$200\Ms$, equal-mass, aligned-spin BBH mergers
(Sec.~\ref{GW190403_like}, Fig.~\ref{fig:diffrate_PSNgap}). These results
suggest that the tandem of massive binary evolution and dynamical interactions
in $\lesssim100$ Myr old YMCs can plausibly produce BBH mergers involving BHs within the PSN gap
and in rates consistent with what is inferred from to-date GW observations.

However, the merger rate density of GW190521-like events from
the present models ($0 - \approx3.8\times10^{-2}\peryg$) is small compared to the models'
overall PSN-gap BBH merger rate density (Table~\ref{tab_rates}, Fig.~\ref{fig:diffrate_PSNgap}).
Also, the model merger rate
density of GW190521-like events fail to reach LVK's upper limit of the same,
as opposed to the comparison for the less massive PSN-gap mergers, and its
average value is several factors smaller than
the LVK's lower limit (see Fig.~\ref{fig:diffrate_PSNgap}).
This means that the present YMC models are rather inefficient in producing GW190521-type
mergers \citep[see also][]{Fragione_2021},
although the upper limit of such events' yield is still consistent with
and within range of the corresponding LVK-estimated rate limits.
Alternative formation channels (see Sec.~\ref{intro} and references therein)
may be more efficient in producing such massive BBH mergers.

Note that the results presented here do not comprise a `full story'
but represent the yield from YMCs.
The modelled clusters represent the most massive young clusters that we observe,
\eg, R136, Westerlund 1. Also, the clusters are evolved up to 300 Myr,
when they are still in their young phase,
dense, and contain a substantial BH population. They would, therefore,
continue to produce late-time GR mergers as they evolve into open clusters,
as seen in Ba21. In the near future, such a cluster model set with a range
of mass, size, and long-term evolution will be investigated.

\begin{acknowledgements}

The author (SB) thanks the reviewer for criticisms and suggestions that have improved the paper.
SB acknowledges support from the Deutsche Forschungsgemeinschaft (DFG; German Research Foundation)
through the individual research grant ``The dynamics of stellar-mass black holes in
dense stellar systems and their role in gravitational-wave generation'' (BA 4281/6-1; PI: S. Banerjee).
SB acknowledges the generous support and efficient system maintenance of the
computing teams at the AIfA and HISKP.
This work and the illustrations presented therein are
greatly benefited by the use of the {\tt Python} packages {\tt NumPy},
{\tt SciPy}, and {\tt Matplotlib}.
This work has been benefited by discussions with Aleksandra Olejak,
Francesco Rizzuto, Chris Belczynski, Giacomo Fragione, Kyle Kremer,
Fabio Antonini, Mark Gieles,
Thorsten Naab, Jeremiah Ostriker, and Rainer Spurzem.
SB has solely performed, managed, and analysed all the N-body computations presented in this work.
SB has done all the coding necessary for this work, prepared the illustrations, and
written the manuscript.

\end{acknowledgements}

\bibliographystyle{aa}
\bibliography{bibliography/biblio.bib}

\appendix

\onecolumn

\section{Direct N-body runs of model star clusters}\label{nbrun}

\begin{ThreePartTable}

\renewcommand*{\arraystretch}{1.5} 
\renewcommand\TPTminimum{\textwidth}

\begin{TableNotes}
\item   {\bf Note:} The columns from left to right give the model
	cluster's ID number, initial mass, $\mcl$, initial half-mass radius, $\rh$,
	initial King concentration parameter, $W_0$,
	metallicity, $Z$, initial overall fraction of primordial binaries, $\fbin$,
	model evolutionary time, $\tevol$, remnant-mass and PPSN/PSN model,
	remnant natal kick model, BH natal spin model,
	number of GR mergers within the cluster, $\nmrgin$, number of GR mergers
	after getting ejected from the cluster, $\nmrgout$.
\end{TableNotes}

\begin{longtable}{>{\stepcounter{rowno}\therowno}rcccclcllrcc}
	\caption{Summary of the new direct N-body evolutionary models of star clusters and their GR-merger yields
	in this work.}\label{tab_nbrun}\\
	\hline
	\hline
	\multicolumn{1}{r}{No.} & \mcl/\Ms     & \rh/pc & $W_0$ & $Z$ & \fbin & \tevol/Myr & remnant model & SN kick & BH spin & \nmrgin & \nmrgout \\
	\hline
	\endfirsthead
        
	\multicolumn{7}{c}%
        {{\bfseries \tablename\ \thetable{} -- continued from previous page}} \\
        \hline
	\hline
	\multicolumn{1}{r}{No.} & \mcl/\Ms     & \rh/pc & $W_0$ & $Z$ & \fbin & \tevol/Myr & remnant model & SN kick & BH spin & \nmrgin & \nmrgout \\
	\hline
	\endhead

	\hline \multicolumn{10}{r}{{Continued on next page}} \\ \hline
        \endfoot

        \hline \hline
	\insertTableNotes 
        \endlastfoot

	&   \footnote{Initial number of stars $N\approx1.28\times10^5$}

	    $7.5\times10^4$ & 2.0       & 7.0 & 0.0002 &  0.05\footnote{The binary fraction is defined
	as $\fbin=2\nbin/N$, $\nbin$ being the total number of binaries and $N$ being
	the total number of members. Note that the initial binary fraction among the O stars, considered
	separately, is $\fobin=1.0$ as opposed to the smaller \emph{overall} binary fraction, $\fbin$. }
								  &  300.0   &   rapid+B16  &  mom. cons.\footnote{Momentum
								  conserving natal kick as in Ba20.}

													  &  FM19\footnote{Vanishing
													  BH natal spins \citep{Fuller_2019a}.}
													             &   0   &   3   \\
	&                   &           &     &        &          &          &              &             &          &   0   &   1   \\
	&                   &           &     &        &          &          &              &             &          &   0   &   1   \\
	&                   &           &     &        &          &          &              &             &          &   2   &   2   \\
	\hline
	&   $7.5\times10^4$ & 2.0       & 7.0 & 0.001  &  0.05    &  300.0   &   rapid+B16  &  mom. cons. &  FM19    &   1   &   3   \\
	&                   &           &     &        &          &          &              &             &          &   2   &   1   \\
	&                   &           &     &        &          &          &              &             &          &   0   &   1   \\
	&                   &           &     &        &          &          &              &             &          &   1   &   0   \\
	\hline
	&   $7.5\times10^4$ & 2.0       & 7.0 & 0.005  &  0.05    &  300.0   &   rapid+B16  &  mom. cons. &  FM19    &   0   &   1   \\
	&                   &           &     &        &          &          &              &             &          &   0   &   0   \\
	&                   &           &     &        &          &          &              &             &          &   0   &   1   \\
	&                   &           &     &        &          &          &              &             &          &   0   &   0   \\
        \hline
	&   $7.5\times10^4$ & 2.0       & 7.0 & 0.01   &  0.05    &  300.0   &   rapid+B16  &  mom. cons. &  FM19    &   0   &   0   \\
	&                   &           &     &        &          &          &              &             &          &   1   &   0   \\
	&                   &           &     &        &          &          &              &             &          &   0   &   0   \\
	&                   &           &     &        &          &          &              &             &          &   0   &   0   \\
        \hline
	&   $7.5\times10^4$ & 2.0       & 7.0 & 0.02   &  0.05    &  300.0   &   rapid+B16  &  mom. cons. &  FM19    &   0   &   1   \\
	&                   &           &     &        &          &          &              &             &          &   0   &   0   \\
	&                   &           &     &        &          &          &              &             &          &   1   &   0   \\
	&                   &           &     &        &          &          &              &             &          &   0   &   0   \\
	\hline
	&   $7.5\times10^4$ & 2.0       & 9.0 & 0.0002 &  0.05    &  300.0   &   rapid+B16  &  mom. cons. &  FM19    &   1   &   2   \\
	&                   &           &     &        &          &          &              &             &          &   2   &   3   \\
	&                   &           &     &        &          &          &              &             &          &   0   &   0   \\
	&                   &           &     &        &          &          &              &             &          &   0   &   1   \\
	\hline
	&   $7.5\times10^4$ & 2.0       & 9.0 & 0.001  &  0.05    &  300.0   &   rapid+B16  &  mom. cons. &  FM19    &   1   &   0   \\
	&                   &           &     &        &          &          &              &             &          &   1   &   1   \\
	&                   &           &     &        &          &          &              &             &          &   0   &   1   \\
	&                   &           &     &        &          &          &              &             &          &   0   &   1   \\
	\hline
	&   $7.5\times10^4$ & 2.0       & 9.0 & 0.005  &  0.05    &  300.0   &   rapid+B16  &  mom. cons. &  FM19    &   2   &   1   \\
	&                   &           &     &        &          &          &              &             &          &   1   &   1   \\
	&                   &           &     &        &          &          &              &             &          &   0   &   0   \\
	&                   &           &     &        &          &          &              &             &          &   1   &   1   \\
	\hline
	&   $7.5\times10^4$ & 2.0       & 9.0 & 0.01   &  0.05    &  300.0   &   rapid+B16  &  mom. cons. &  FM19    &   0   &   0   \\
	&                   &           &     &        &          &          &              &             &          &   1   &   0   \\
	&                   &           &     &        &          &          &              &             &          &   0   &   0   \\
	&                   &           &     &        &          &          &              &             &          &   1   &   0   \\
	\hline
	&   $7.5\times10^4$ & 2.0       & 9.0 & 0.02   &  0.05    &  300.0   &   rapid+B16  &  mom. cons. &  FM19    &   0   &   0   \\
	&                   &           &     &        &          &          &              &             &          &   0   &   0   \\
	&                   &           &     &        &          &          &              &             &          &   0   &   0   \\
	&                   &           &     &        &          &          &              &             &          &   0   &   0   \\
\end{longtable}
\end{ThreePartTable}

\begin{figure*}
\centering
\includegraphics[width = 18.5 cm, angle=0.0]{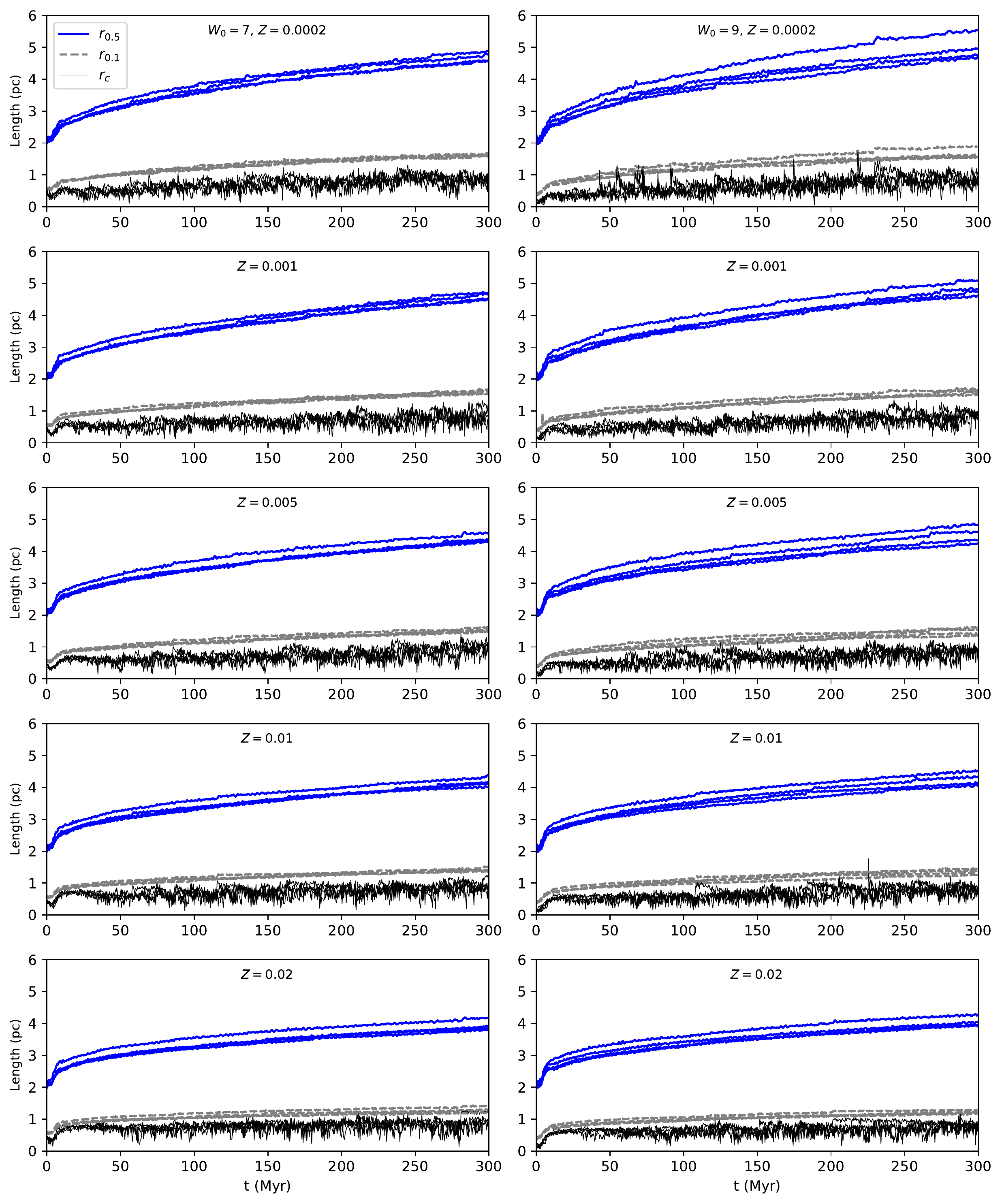}
\caption{The time evolution of the core radius, $r_c$, 10\% Lagrangian radius, $r_{0.1}$,
	and 50\% Lagrangian radius (half mass radius), $r_{0.5}$, of all the computed
	cluster models in Table~\ref{tab_nbrun}. The panels in the left (right)
	column correspond to the models with $W_0=7$ ($W_0=9$) with $Z$
	as indicated in each panel's title. $r_c(t)$, $r_{0.1}(t)$, and $r_{0.5}(t)$
	(legend) are plotted for each model separately.}
\label{fig:sizeevol}
\end{figure*}

\twocolumn

\end{document}